\documentclass{webofc}

\usepackage[varg]{txfonts}   
\usepackage{hyperref}
\usepackage{url}
\hypersetup{colorlinks=true,citecolor=blue,urlcolor=blue,linkcolor=blue}
\begin{document}
\title{Data discovery, analysis and reproducibility in Virtual \newline Research Environments}
%
%

\author{\firstname{Enrique} \lastname{Garc\'ia Garc\'ia}\inst{1}\fnsep\thanks{\email{enrique.garcia.garcia@cern.ch}} \and
        \firstname{Giovanni} \lastname{Guerrieri}\inst{1} \and
        \firstname{Rub\'en} \lastname{P\'erez Mercado}\inst{1} \and
        \firstname{Michael Ryan} \lastname{Zengel}\inst{1} \and
        \firstname{Georgy} \lastname{Skorobogatov}\inst{2,3} \and
        \firstname{Hugo} \lastname{Gonz\'alez Labrador}\inst{1} \and
        \firstname{Xavier} \lastname{Espinal Curull}\inst{1}
}

\institute{European Organization for Nuclear Research (CERN), 1211 Meyrin, Switzerland
\and
           Departament de F\'isica Qu\`antica i Astrof\'isica (FQA), Universitat de Barcelona (UB), Mart\'i i Franqu\'es, 1, 08028 Barcelona, Spain
\and
           Institut de Ci\`encies del Cosmos (ICCUB), Universitat de Barcelona (UB), Mart\'i i Franqu\'es, 1, 08028 Barcelona, Spain
          }

\abstract{
During the ESCAPE project, a pilot analysis facility was developed with a bottom-up approach, in collaboration with all the project partners. As a result, the CERN Virtual Research Environment (VRE) initiative proposes a workspace that facilitates data access from the ESCAPE Data Lake, managed by Rucio, a data management framework, and supports interactive analysis via Jupyter notebooks. The facility offers custom software stacks and access to CVMFS, and connects to local data processing resources through REANA, an open-source framework developed by CERN IT for reanalysis and reproducibility. The CERN VRE deploys an instance of REANA, allowing users to utilise its features together with the analysis facility's services.

Integrating heterogeneous services with a unified interface significantly eases the user experience. Furthermore, in line with the ESCAPE Open Collaboration, the development of open source tools that can be leveraged by different physics communities with similar analysis strategies, laying the foundation of common lifecycle analysis practices. Therefore, in order to foster accessibility, as well as interactivity, reproducibility and data preservation to more complex infrastructure services, the development of user-friendly middleware should be prioritized. 

This contribution focuses on the connection of REANA and Zenodo to the CERN Virtual Research Environment's interface through Jupyter extensions. The development of these extensions makes it possible to use the Virtual Research Environment as a single workspace to enhance the lifecycle of research analysis: from data discovery and data access, through interactive analysis and offload to computing resources, to reproducibility and preservation of results.
}
\maketitle
\section{Context and introduction}
\label{intro}

The ESCAPE\footnote{European Science Cluster of Astronomy \& Particle Physics ESFRI Research Infrastructures} project~\cite{proj-escape} was a European initiative that began in 2019, funded by the European Union’s Horizon 2020 research and innovation program. It brought together various partners from the particle physics, nuclear physics, and astrophysics communities to tackle common challenges and tailor shared solutions. The focus areas identified by ESCAPE can be summarized by the work packages existing within the project: Data Infrastructure for Open Science, Open-source Scientific Software and Service Repository, International Virtual Observatory Alliance, Science Analysis Platforms, and Citizen Science.

On January 2023, all the partners of the project signed a collaboration agreement and the project became an Open Collaboration~\cite{escape-ca}.

\section{The Virtual Research Environment}
\label{sec-cern-vre}

The Virtual Research Environment (VRE), is an analysis facility prototype built to fulfill the ESCAPE scientists' needs, that evolved conceptually and technologically throughout the project's duration. It was originally designed as the place where data-related activities would take place (see, for example, DAC21, the ESCAPE Data and Analysis Challenge on November 2021~\cite{DAC21}) as well as the entry point for other communities to test and try Rucio~\cite{paper-rucio}, a scientific data management solution developed at CERN (see Section~\ref{subsec-underneath-ser}). It served as well as the entry point to the ESCAPE Data Lake~\cite{escape-datalake, escape-datalake2}. Throughout the ESCAPE project, the VRE also provisioned the members of the EOSC Future project~\cite{eosc-future} with a common platform on which they could carry out all the stages of their scientific studies~\cite{eosc-sp, sp-repo}. These interactions allowed a bottom-up approach in the conception of the VRE, addressing ESCAPE users' needs while developing common tools and methodologies, complying with the FAIR principles~\cite{fair} from the outset.

To date, the VRE is an open-source multi-purpose platform that showcases one of the many existing analysis facilities concepts. While serving as proof of concept of certain ESCAPE outcomes, it acts as a middleware interface to the different services and computing resources used to execute an analysis.

\subsection{The VRE Architecture}
\label{subsec-vre-arch}

The Virtual Research Environment is deployed on a Kubernetes cluster (k8s)~\cite{k8s} leveraging CERN's OpenStack cloud infrastructure~\cite{openstack}. The cluster orchestrates the deployment of the services and connects the different applications with CERN resources (storage, network, etc.). Currently, the cluster runs on version \texttt{v1.29.2}. 

\begin{figure}[h]
\centering
\includegraphics[width=0.95\textwidth,clip]{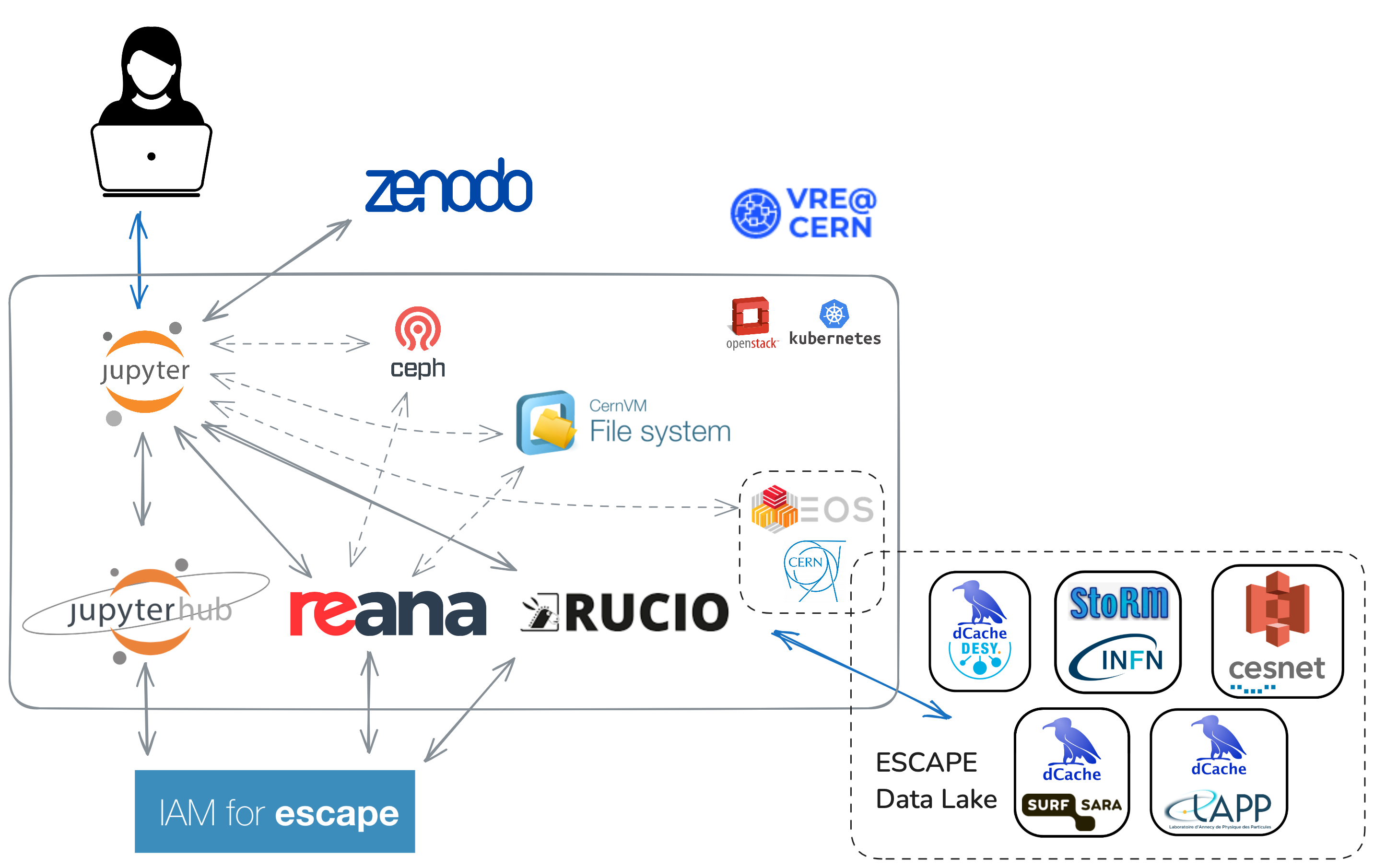}
\caption{Schematic architecture of the CERN Virtual Research Environment and the interaction between its components.}
\label{vre-architecture}       
\end{figure}

The VRE services are instanced using Helm~\cite{helm}, whose charts are kept on a GitHub public repository~\cite{repo-vre} for version control. Continuous delivery is achieved via the Flux~\cite{flux} system and Sealed Secrets~\cite{sealed-secrets} is used for secret management and encryption. Figure~\ref{vre-architecture} shows a general view of the platform's architecture and a schematic interaction between the different applications. The VRE components are described in Section~\ref{subsec-underneath-ser} and their current deployed versions are summarised in Table~\ref{tab-1}.

\begin{table}[h]
\centering
\caption{Current versions of the VRE deployments and components}
\label{tab-1}       
\begin{tabular}{|ccc|ccc|}
\hline
Rucio & Reana & JupyterHub & K8s cluster & CERN magnum & Sealed-Secrets   \\ \hline
34.6.0 & 0.9.3 & 4.1.5 & 1.29.2 & 0.15.2 & 0.27.0 \\ \hline
\end{tabular}
\end{table}

\subsection{The VRE core services}
\label{subsec-underneath-ser}

This section briefly summarises the VRE core components, focusing on their latest developments. Further details about the main applications of the platform can be found on the VRE documentation page~\cite{vre-docs} and in previous publications~\cite{vre-chep-22}.

The VRE Authentication and Authorization Infrastructure (AAI) relies on an  INDIGO IAM (Identity and Access Management) instance created for the ESCAPE project~\cite{iam}. IAM is compatible both with X.509 certificates (currently being deprecated) and OIDC token-based authorization, based on WLCG JWT profiles~\cite{wlcg-jwt-profile}. Contrary to all the following services, the ESCAPE IAM instance relies on a third party deployment.

A JupyterHub instance~\cite{jupyterhub} provisions JupyterLab~\cite{jupyterlab} sessions to any registered user. Pre-loaded custom software environments can be spawned by installing the stack on top of Jupyter-based Docker images~\cite{docker}, available when logging into the VRE. Also, CVMFS~\cite{cvmfs} distributions are programmatically accessible from any user's session, as the filesystem is mounted on the cluster. A shared CEPHFS volume~\cite{cephfs} is available for users' storage, as well as an EOS instance~\cite{eos}, currently employed as one of the storage technologies within the ESCAPE Rucio service.

The ESCAPE Rucio instance~\cite{escape-datalake, vre-chep-22} provides a data management solution framework for the collaboration's users and partners. Rucio connects the storage elements of different data centers, managing the data transfers between them on users' needs. Rucio uses FTS3~\cite{fts} and GFAL2~\cite{gfal} to enable a transparent and seamless interaction between different storage technologies and file transfer protocols. A CERN EOS instance (hosted at \texttt{eospilot.cern.ch}) serves as a Rucio Storage Element as part of the ESCAPE Data Lake infrastructure, together with various other data centers~\cite{vre-chep-22}.

A Reana instance~\cite{reana} allows users to leverage the clusters' computing resources. The advantages of provisioning Reana within the cluster are the flexibility to use various computational workflow engines (CWL, snakemake, and yadage), as well as the reproducibility that is implicit with Reana. The deployment uses a CEPHFS volume to run and store any Reana result. The VRE-Reana instances is hosted on a dedicated URL~\cite{reana-vre}. Reana also allows the possibility to offload heavy analysis workflows to different computational backends, like HTCondor, Slurm or Kubernetes; the latter is currently employed by the VRE.

Each of the services described can be used independently; however, the VRE makes them available through a unique entry point. Daily jobs run in the cluster's background, synchronizing user identities and enabling users to interact with all the services through a single ESCAPE IAM login. 

\subsection{End-to-end scientific workflows}

\begin{figure}[h]
\centering
\includegraphics[width=0.9\textwidth, clip]{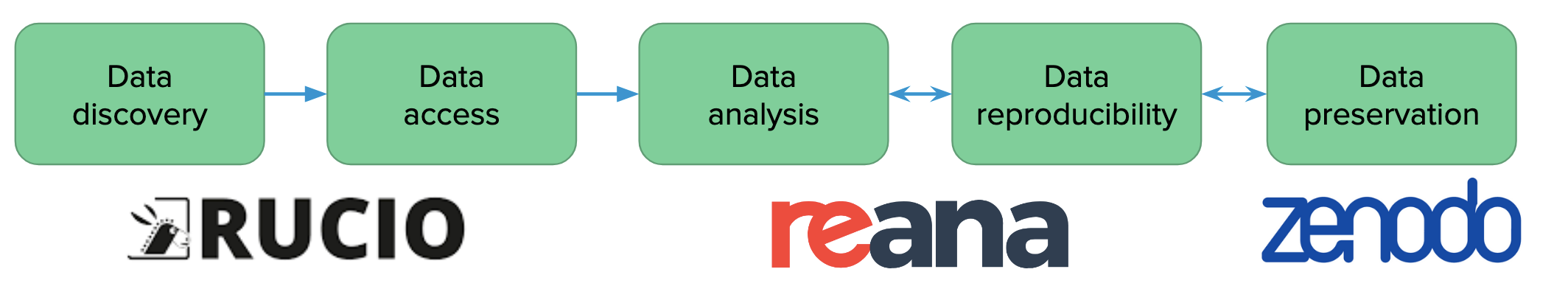}
\caption{Schema representing the scientific workflow stages and the VRE services that cover them.}
\label{img-stages-analysis}       
\end{figure}

Every scientific workflow or analysis is unique and depends on the field, experiment, and researcher. Within ESCAPE, a few common stages have been identified and shown in Figure~\ref{img-stages-analysis}: data discovery, data access, analysis of data, reproducibility results, and data preservation.

During 2024, two additional JupyterLab extensions were developed to seamlessly connect the underlying VRE services interactively. This choice was motivated by the positive experience with the Rucio JupyterLab extension (see Section~\ref{subsec-rucio-jlab}), and by the commitment to lower the barrier to entry for the VRE services, simplifying the user experience, and aggregating all the stages of a workflow on a single application.


\subsubsection{Rucio JupyterLab extension}
\label{subsec-rucio-jlab}

The Rucio JupyterLab extension~\cite{rucio-jlab} was developed as part of a Google Summer of Code internship~\cite{gsoc} in 2022. 

\begin{figure}[h]
\centering
\includegraphics[width=0.65\textwidth, clip]{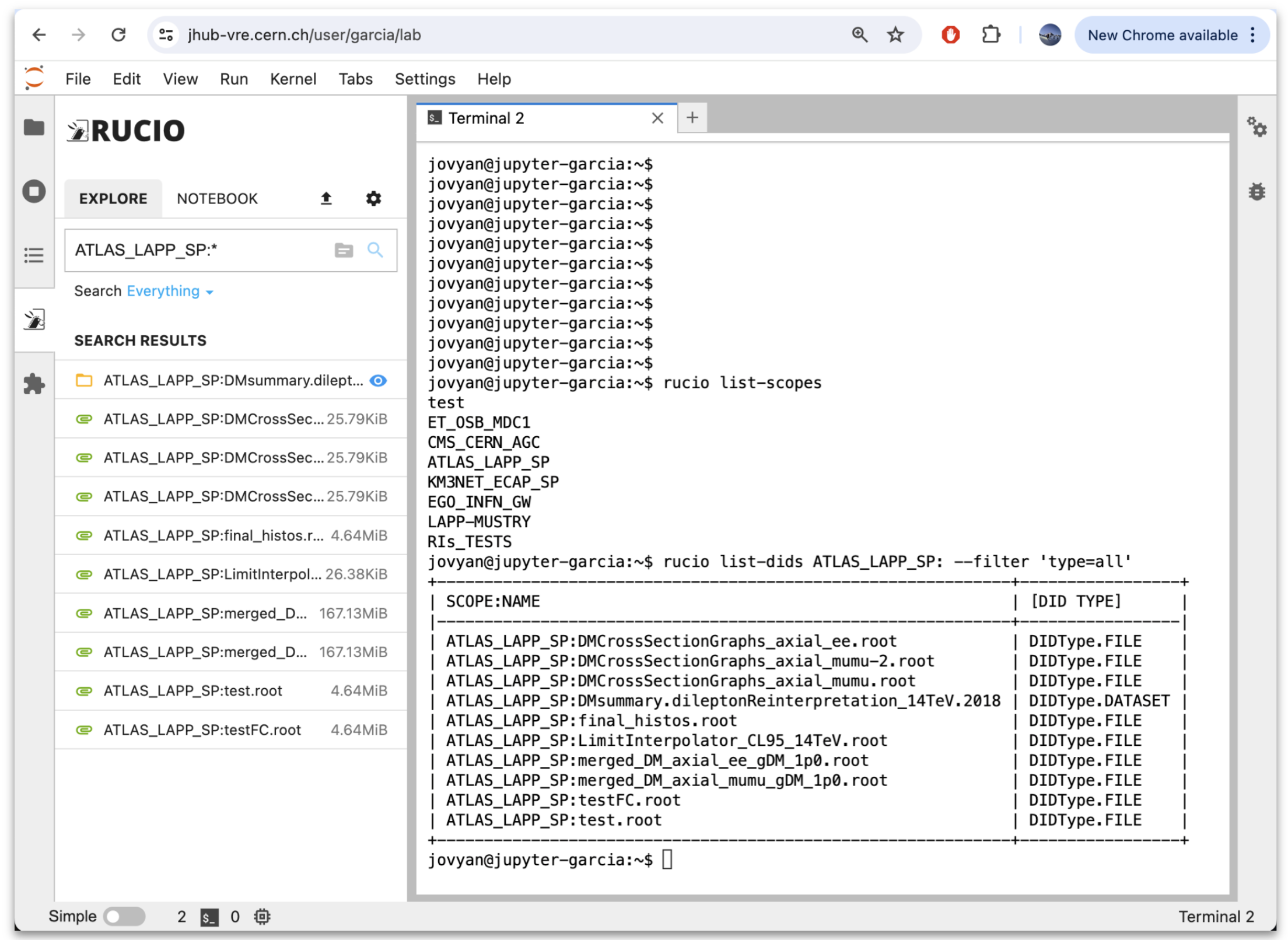}
\caption{Example of usage of the Rucio JupyterLab extension browsing a specific Rucio data set (left) vs. its equivalent CLI commands (right).}
\label{img-rucio-jlab}       
\end{figure}

The main purpose of the extension is to ease the access and discoverability of data from a Data Lake by bridging the gap between the Rucio CLI and the Jupyter-based analysis platform (see Figure~\ref{img-rucio-jlab}). If a Rucio storage element is FUSE mounted on the user's session, the extension allows the injection of files present on this volume to a Jupyter Notebook as variables. Also, the plugin allows triggering the replication of data to the attached volume if the data is not present.

\subsubsection{Reana JupyterLab extension}
\label{subsec-reana-jlab}

The Reana extension~\cite{reana-jlab} was developed during the summer of 2024 as part of an Openlab CERN summer student programme~\cite{ss-programme}. Like the Rucio extension, this plugin was developed to ease the interaction of users with the Reana instance available on the VRE cluster, covering the data analysis and data reproducibility stages of a workflow lifecycle (see Figure~\ref{img-stages-analysis}) from the VRE JupyterLab interface.

After the user's authentication, the JupyterLab plugin interacts with the Reana-Server~\cite{reana-server} REST API endpoints using the Reana-clients library~\cite{reana-client}.

\begin{figure}[h]
\centering
\includegraphics[width=0.95\textwidth,clip]{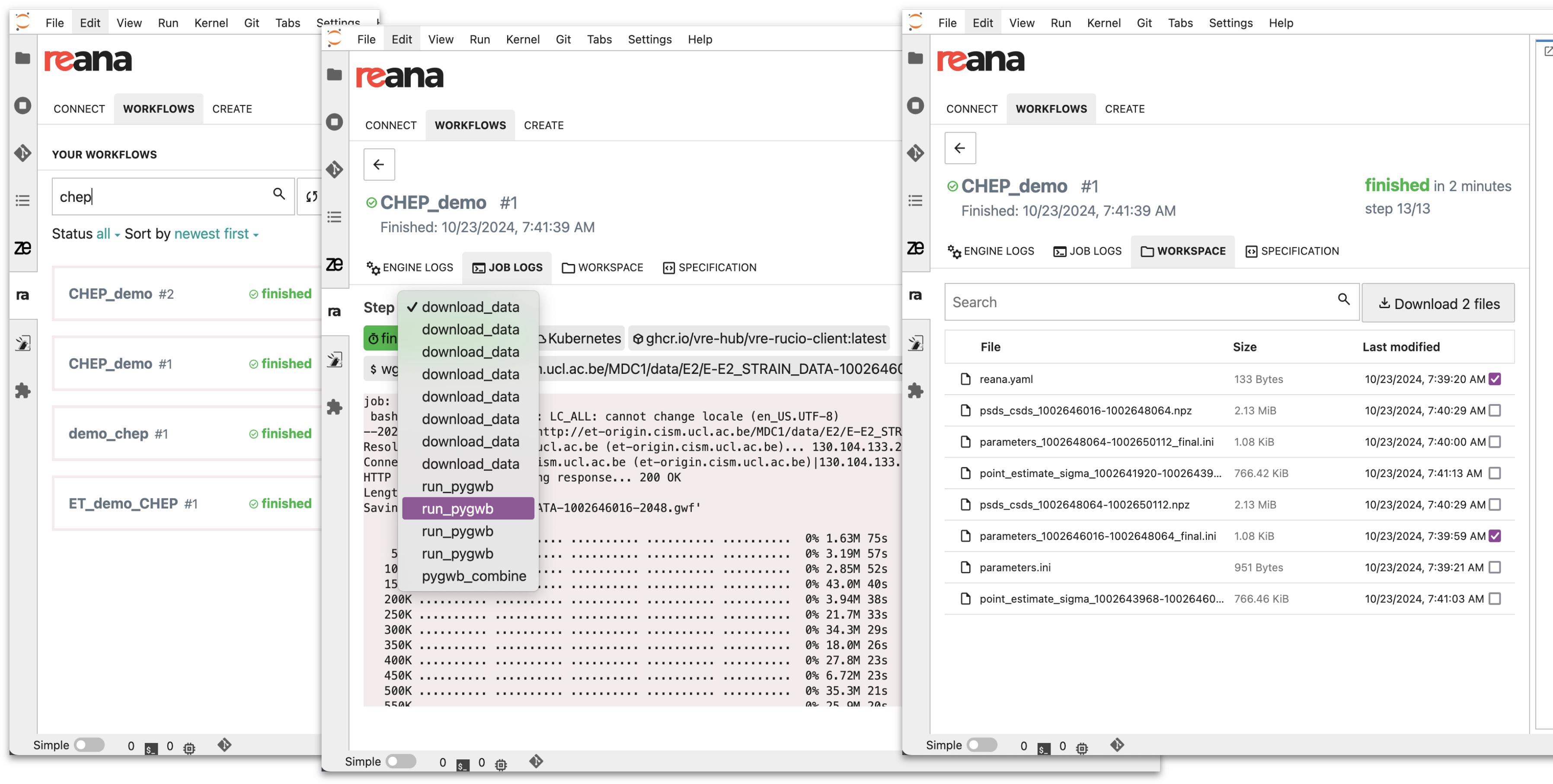}
\caption{Various JupyterLab snapshots showing the displays of the Reana JupyterLab extension.}
\label{img-reana-jlab}       
\end{figure}

The main functionalities that the Reana plugin offers are very similar to those available on a Reana user interface (visit \cite{reana-vre} in the case of the Reana-VRE instance); it shows the running status of a workflow and allows browsing and interacting with previous executed workflows (see Figure~\ref{img-reana-jlab}). The benefits that the extension brings are i), the possibility of starting Reana workflows from the plugin and ii), it allows downloading the results of a successful workflow to the storage of the user's Jupyter session, enabling the interaction workflow results.

\subsubsection{Zenodo JupyterLab extension}
\label{subsec-zenodo-jlab}

The Zenodo extension~\cite{zenodo-jlab} was developed during the summer of 2024 as part of a CERN summer student programme~\cite{ss-programme}. It allows interaction with Zenodo~\cite{zenodo}, a multi-purpose repository developed at CERN that provides a unique DOI (Digital Object Identifier) to any digital record uploaded to the platform. This extension allows the VRE users to close the scientific life cycle by preserving results or uploading any scientific digital asset, to a trusted and known repository.

Similarly as above, the extension exchanges REST API calls of the user's Jupyter session server with the Zenodo REST API~\cite{zenodo-api}. The client package used for this exchange is the eOSSR library~\cite{eossr}, a package developed in the context of the ESCAPE OSSR (Open-source Scientific Software and Service Repository)~\cite{ossr}). The ESCAPE repository is hosted on a Zenodo community named \texttt{escape2020}~\cite{ossr-zenodo-community} and is used to preserve all the ESCAPE digital products (documents, scientific results, software, etc.).

\begin{figure}[h]
\centering
\includegraphics[width=0.99\textwidth,clip]{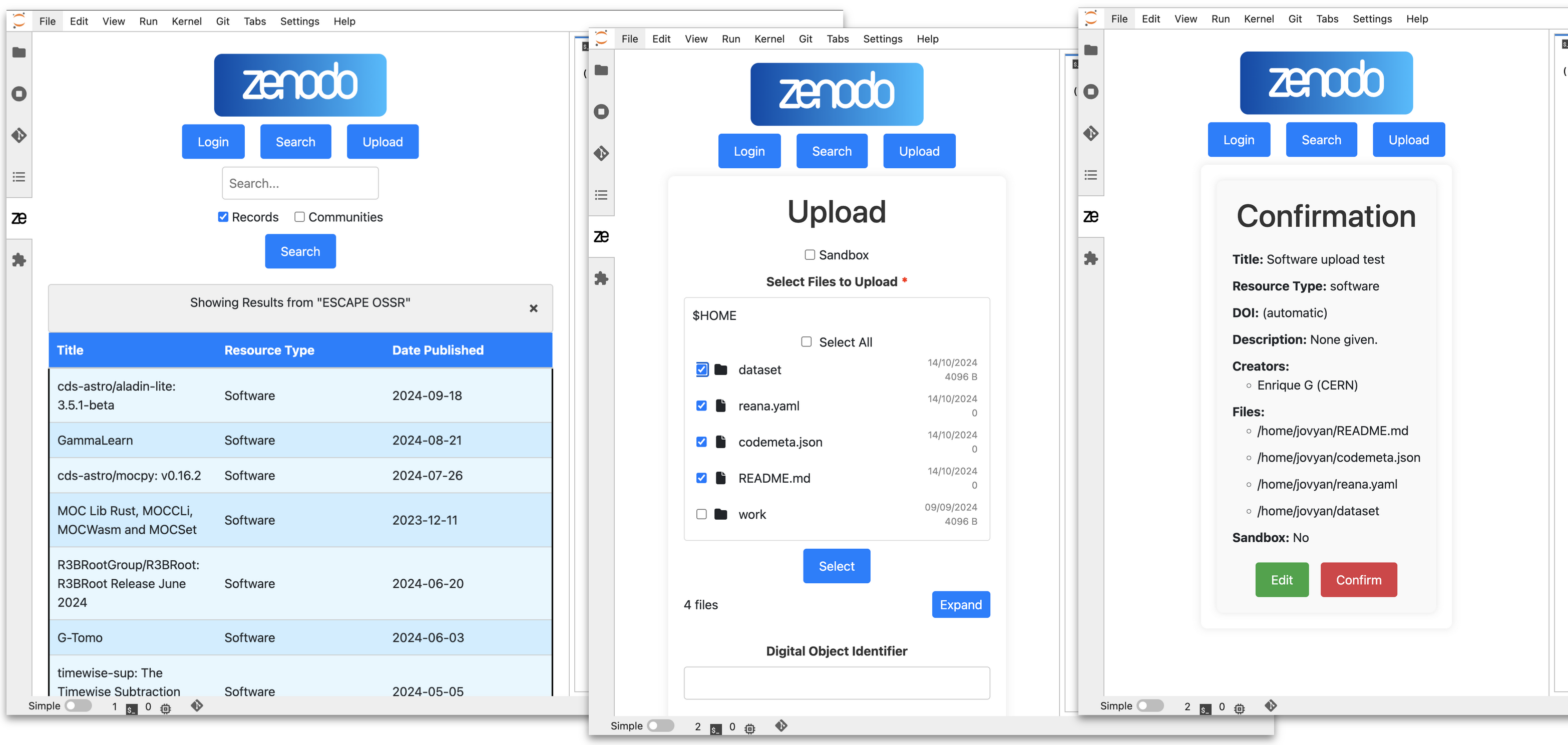}
\caption{JupyterLab snapshots showing the different displays of the Zenodo JupyterLab extension.}
\label{img-zenodo-jlab}       
\end{figure}

\subsection{Reproducible demo analysis}

A recorded demo of an end-to-end scientific workflow using the VRE is available online~\cite{vre-demo}. The source code and instructions on how to run the analysis are publicly accessible~\cite{sp-repo, repo-pygwb}, as well as the instructions to logging into the VRE~\cite{vre-docs}.

\section{Summary and next steps}

The Virtual Research Environment is an analysis facility prototype developed during the ESCAPE project. It allows users to perform all the stages of scientific analysis: access and explore data through a Data Lake, (re)analyze collections of data, offload larger analyses to a computing backend, and preserve the results under a unique DOI. 

Deployed within CERN resources, the VRE simplifies the interaction between the Rucio, the Reana, and the Zenodo services, connecting them via JupyterLab extensions on a Jupyter framework.

The platform has served as a demonstrator for certain solutions developed during the ESCAPE project, as shown by the adoption of the Rucio framework in other large scientific communities (SKA(O), CTA(O), KM3NeT, and the Einstein Telescope~\cite{madden}). In certain cases, it has also helped laying the foundations for similar analysis facilities~\cite{madden}. Finally, the VRE has also been used as a platform for educational and Open Science activities~\cite{workshop-cuoco}.

\paragraph{Acknowledgements. \rm{This work was partially supported by MCIN with funding from European Union NextGenerationEU(PRTR-C17.I1) and by Generalitat de Catalunya.}}

%
%
%

\end{document}